

\input lanlmac


\newcount\figno
\figno=0
\def\fig#1#2#3{
\par\begingroup\parindent=0pt\leftskip=1cm\rightskip=1cm\parindent=0pt
\baselineskip=11pt
\global\advance\figno by 1
\midinsert
\epsfxsize=#3
\centerline{\epsfbox{#2}}
\vskip 12pt
\centerline{{\bf Fig.\the\figno~} #1}\par
\endinsert\endgroup\par
}
\def\figlabel#1{\xdef#1{\the\figno}}

\def\N{{\cal N}}

\def\th{\theta}

\def\cob{\delta}
\def\ep{\epsilon}

\def\hf{{1\over 2}}
\def\qu{{1\over 4}}

\def\o{\over}

\def\bra{\langle}
\def\ket{\rangle}
\def\lf{\left}
\def\ri{\right}
\def\riya{\rightarrow}

\def\h#1{\widehat{#1}}

\def\al{\alpha}

\def\tens{\otimes}
\def\Om{\Omega}
\def\dag{\dagger}
\def\rt#1{\sqrt{#1}}
\def\st{\star}

\def\sitarel#1#2{\mathrel{\mathop{\kern0pt #1}\limits_{#2}}}

\def\V{V_{00}^{rr}}

\lref\Mpri{
B.~Feng, Y.~H.~He and N.~Moeller,
``The spectrum of the Neumann matrix with zero modes,''
[hep-th/0202176].
}

\lref\Schnabl{
M. Schnabl,
``Wedge states in string field theory,''
[hep-th/0201095].
}
\lref\KOkin{
K.~Okuyama,
``Ghost kinetic operator of vacuum string field theory,''
[hep-th/0201015].
}
\lref\HKcheck{
L.~Rastelli, A.~Sen and B.~Zwiebach,
``A note on a proposal for the tachyon state in vacuum string field  theory,''
[hep-th/0111153]\semi
R.~Rashkov and K.~S.~Viswanathan,
``A Note on the Tachyon State in Vacuum String Field Theory,''
[hep-th/0112202].
}
\lref\HI{
A.~Hashimoto and N.~Itzhaki,
``Observables of string field theory,''
[hep-th/0111092].
}
\lref\KishiOh{
I.~Kishimoto and K.~Ohmori,
``CFT description of identity string field: 
Toward derivation of the VSFT  action,''
[hep-th/0112169].
}
\lref\siegel{
K.~Okuyama,
``Siegel gauge in vacuum string field theory,''
[hep-th/0111087].
}
\lref\spec{
L.~Rastelli, A.~Sen and B.~Zwiebach,
``Star algebra spectroscopy,''
[hep-th/0111281].
}
\lref\GRSZ{
D.~Gaiotto, L.~Rastelli, A.~Sen and B.~Zwiebach,
``Ghost Structure and Closed Strings in Vacuum String Field Theory,''
[hep-th/0111129].
}

\lref\Witten{
E. Witten,
``Noncommutative Geometry And String Field Theory,''
Nucl.\ Phys.\  {\bf B268} (1986) 253.
}
\lref\HorowitzYZ{
G.~T.~Horowitz and A.~Strominger,
``Translations As Inner Derivations And Associativity Anomalies In Open
String Field Theory,''
Phys.\ Lett.\  {\bf B185} (1987) 45.
}
\lref\LPP{
A. LeClair, M. E. Peskin and C. R. Preitschopf,
``String Field Theory on the Conformal Plane. 1. Kinematical Principles,''
Nucl. Phys. {\bf B317} (1989) 411\semi
``String Field Theory on the Conformal Plane. 2. Generalized Gluing,''
Nucl. Phys. {\bf B317} (1989) 464.
}
\lref\GrossJ{
D. J. Gross and A. Jevicki,
``Operator Formulation of Interacting String Field Theory,''
Nucl.\ Phys.\ {\bf B283} (1987) 1;
``Operator Formulation of Interacting String Field Theory (II),''
Nucl.\ Phys.\ {\bf B287} (1987) 225.
}

\lref\RZ{
L. Rastelli and B. Zwiebach,
``Tachyon potentials, star products and universality,''
JHEP {\bf 0109}, 038 (2001)
[hep-th/0006240].
}

\lref\VSFT{
L.~Rastelli, A.~Sen and B.~Zwiebach,
``String Field Theory Around the Tachyon Vacuum,''
[hep-th/0012251];
``Vacuum string field theory,''
[hep-th/0106010].
}

\lref\KP{
V. A. Kostelecky and R. Potting,
``Analytical construction of a nonperturbative vacuum
for the open bosonic string,''
Phys. Rev. {\bf D63} (2001) 046007 [hep-th/0008252].
}

\lref\Ohmori{
K.~Ohmori,
``A review on tachyon condensation in open string field theories,''
[hep-th/0102085].
}

\lref\RSZ{
L. Rastelli, A. Sen and B. Zwiebach,
``Half-strings, Projectors, and Multiple D-branes
in Vacuum String Field Theory,''
[hep-th/0105058];
}
\lref\BCFT{
L. Rastelli, A. Sen and B. Zwiebach,
``Boundary CFT construction of D-branes in vacuum string field theory,''
[hep-th/0105168];
}
\lref\RSZclas{
L. Rastelli, A. Sen and B. Zwiebach,
``Classical Solutions in String Field Theory Around the Tachyon Vacuum,''
[hep-th/0102112];
}
\lref\GT{
D. J. Gross and W. Taylor,
``Split string field theory I,'' JHEP {\bf 0108}, 009 (2001),
[hep-th/0105059];
``Split string field theory. II,''
JHEP {\bf 0108}, 010 (2001)
[hep-th/0106036].
}
\lref\KO{
T.~Kawano and K.~Okuyama,
``Open string fields as matrices,''
JHEP {\bf 0106}, 061 (2001)
[hep-th/0105129].
}
\lref\JD{
J.~R.~David,
``Excitations on wedge states and on the sliver,''
JHEP {\bf 0107}, 024 (2001)
[hep-th/0105184].
}
\lref\Mu{
P.~Mukhopadhyay,
``Oscillator representation of the BCFT construction of D-branes in  vacuum string field theory,''
[hep-th/0110136].
}
\lref\HK{
H.~Hata and T.~Kawano,
``Open string states around a classical 
solution in vacuum string field  theory,''
JHEP {\bf 0111}, 038 (2001)
[hep-th/0108150].
}
\lref\matsuo{
Y.~Matsuo,
``BCFT and sliver state,''
Phys.\ Lett.\ B {\bf 513}, 195 (2001)
[hep-th/0105175];
``Identity projector and D-brane in string field theory,''
Phys.\ Lett.\ B {\bf 514}, 407 (2001)
[hep-th/0106027];
``Projection operators and D-branes in purely 
cubic open string field  theory,''
Mod.\ Phys.\ Lett.\ A {\bf 16}, 1811 (2001)
[hep-th/0107007].
}
\lref\Kishimoto{
I.~Kishimoto,
``Some properties of string field algebra,''
JHEP {\bf 0112}, 007 (2001)
[hep-th/0110124].
}
\lref\HataMoriyama{
H.~Hata and S.~Moriyama,
``Observables as Twist Anomaly in Vacuum String Field Theory,''
[hep-th/0111034].
}
\lref\FO{
K.~Furuuchi and K.~Okuyama,
``Comma vertex and string field algebra,''
JHEP {\bf 0109}, 035 (2001)
[hep-th/0107101].
}
\lref\Moeller{
N.~Moeller,
``Some exact results on the matter star-product 
in the half-string  formalism,''
[hep-th/0110204].
}
\lref\moore{
G. Moore and W. Taylor,
``The singular geometry of the sliver,''
[hep-th/0111069].
}

\Title{             
                                             \vbox{\hbox{EFI-02-60}
                                             \hbox{hep-th/0201136}}}
{\vbox{
\centerline{Ratio of Tensions from Vacuum String Field Theory}
}}

\vskip .2in

\centerline{Kazumi Okuyama}

\vskip .2in

\centerline{ Enrico Fermi Institute, University of Chicago} 
\centerline{ 5640 S. Ellis Ave., Chicago IL 60637, USA}
\centerline{\tt kazumi@theory.uchicago.edu}

\vskip 3cm
\noindent

We show analytically that the ratio of the norm of sliver states
agrees with the ratio of D-brane tensions. 
We find that the correct ratio appears as a twist anomaly. 
 
\Date{January 2002}

\vfill
\vfill

\newsec{Introduction}
Vacuum String Field Theory (VSFT) is proposed \VSFT\ as a
bosonic open string field theory around the closed string vacuum.
(See \refs{\RSZclas\RSZ\GT\KO\BCFT\matsuo\JD\FO\HK\Kishimoto\Mu\Moeller
\HataMoriyama\moore\siegel\HI\GRSZ\HKcheck\spec\KishiOh\KOkin{--}\Schnabl}
for related papers.) 
One of the interesting problems in VSFT is to
reconstruct the unstable vacua describing 
D-branes as classical solutions of VSFT and 
extract the physical quantities such as the ratio of tensions 
\refs{\RSZclas,\RSZ}
and the ratio of the potential height to the tension
\refs{\HK,\HataMoriyama,\HKcheck}. 

Since the kinetic operator of VSFT is supposed to be purely ghost,
the equation of motion
\eqn\eom{
Q\Psi+\Psi\st\Psi=0
}
has a matter-ghost factorized solution \RSZclas:
\eqn\factsol{
\Psi=\Psi_m\tens\Psi_g,
}
where the matter part $\Psi_m$  is given by
a projection
\eqn\projPsi{
\Psi_m\st_m\Psi_m=\Psi_m,\quad Q\Psi_g+\Psi_g\st_g\Psi_g=0.
}
Assuming that the ghost part $\Psi_g$ are common to all the solutions,
the ratio of the action is given by the ratio of the norm of projections:
\eqn\actratio{
{S[\Psi']\o S[\Psi]}={\bra \Psi'_m|\Psi'_m\ket\o\bra \Psi_m|\Psi_m\ket}.
}
In \RSZclas, Rastelli, Sen and Zwiebach obtained 
a closed form for this ratio
and checked numerically that this ratio
reproduces the ratio of D-brane tensions.  
In this paper, we calculate this ratio analytically 
in the oscillator representation of string fields
(see \BCFT\ for the computation of 
this ratio from the BCFT approach).

Naively, the norm of projection is vanishing \refs{\RSZclas,\GRSZ,\KOkin},
so the ratio \actratio\ has an ill-defined expression ${0\o0}$.
By carefully regularizing the ratio of determinants appearing in the norm of
projection, we find that the correct ratio appears as a twist anomaly
as discussed in \HataMoriyama.

This paper is organized as follows:
In section 2, we review the conjecture of \RSZclas.
In section 3, we summarize the properties of Neumann coefficients
which are used in the calculation.
In section 4, we calculate the ratio of the norm of projections
and show that this agrees with the expected ratio of D-brane tensions. 
Section 5 is devoted to discussions.
In Appendix A, we give a derivation of $\bra k|v_e\ket$ and $\bra k|v_o\ket$.
Appendix B discusses the integral representation of the identities obtained
in \HataMoriyama.

\newsec{RSZ Conjecture}
In this section, we review the statement of the conjecture in \RSZclas.
Since the 3-string vertex has a factorized form 
with respect to the spacetime directions,
here we focus on one spatial direction. 
The matter part of the
3-string vertex in the $\al'=1$ convention is written as 
\refs{\GrossJ,\RSZclas}
\eqn\Vth{\eqalign{
|V_3\ket&=\int\prod_{r=1}^3dp_{(r)}\cob\lf(\sum_{s=1}^3p_{(s)}\ri)
\exp\lf(\sum_{r,s=1}^3-\hf \sum_{n,m=1}^{\infty}a_n^{(r)\dag}
V^{rs}_{nm}a_m^{(s)\dag}-\sum_{n=1}^{\infty}
p_{(r)}V^{rs}_{0n}a_n^{(s)\dag}\ri.\cr
&\lf.\hskip 57mm -\hf\sum_{r=1}^3p_{(r)}\V p_{(r)}\ri)|0,p\ket_{123} \cr
&={(2\pi b^3)^{\qu}\o\rt{3}\lf(\V+{b\o2}\ri)}
\exp\lf(-\hf\sum_{r,s=1}^3\sum_{n,m=0}^{\infty}
a_n^{(r)\dag}V'^{rs}_{nm}a_m^{(s)\dag}\ri)|\Om_b\ket_{123}.
}}
Here  $|\Om_b\ket$ is the vacuum for zero-mode oscillator $a_0$ defined by
\eqn\Omb{
a_0|\Om_b\ket=0,\quad a_0={\rt{b}\o2}\h{p}-{i\o\rt{b}}\h{x},
}
where $b$ is a numerical parameter.
We can construct two types of projections: one is the sliver $|\Xi\ket$ 
\refs{\KP,\RSZclas}
and the other is the projection $|\Xi'\ket$ describing the lump solution.
The explicit forms of $|\Xi\ket$ and $|\Xi'\ket$ are given by
\eqn\sliver{\eqalign{
|\Xi\ket&=\det(1-M)^{\hf}(1+T)^{\hf}
\exp\lf(-\hf a^{\dag}CT a^{\dag}\ri)|0\ket, \cr
|\Xi'\ket&={\rt{3}\lf(\V+{b\o2}\ri)\o(2\pi b^3)^{\qu}}
\det(1-M')^{\hf}(1+T')^{\hf}
\exp\lf(-\hf a^{\dag}C'T' a^{\dag}\ri)|\Om_b\ket.
}}
The various matrices appearing in 
$|\Xi\ket$ and $|\Xi'\ket$ are defined by
\eqn\defMMpri{\eqalign{
&M=CV^{11},\quad M'=C'V'^{11}, \cr
&C_{nm}=(-1)^n\cob_{nm}\quad(n,m\geq 1),\qquad 
C'_{nm}=(-1)^n\cob_{nm}\quad (n,m\geq 0), \cr
&T={1\o2M}\lf(1+M-\rt{(1-M)(1+3M)}\ri),~~
T'={1\o2M'}\lf(1+M'-\rt{(1-M')(1+3M')}\ri).
}}
Note that the indices of primed matrices run from $0$ to $\infty$,
whereas the indices of unprimed matrices run from $1$ to $\infty$.
$C$ is called a twist matrix, and $M$ commutes with $C$.

In \RSZclas, Rastelli, Sen and Zwiebach conjectured that the
ratio $R$ of the norm of $|\Xi\ket$ and $|\Xi'\ket$ given by
\eqn\Ratio{\eqalign{
R&={\bra 0|0\ket\o\bra \Xi|\Xi\ket}{\bra \Xi'|\Xi'\ket\o\bra\Om_b|\Om_b\ket}\cr
&={3\lf(\V+{b\o2}\ri)^2\o\rt{2\pi b^3}}
{\det(1-M')^{{3\o4}}(1+3M')^{\qu}\o\det(1-M)^{{3\o4}}(1+3M)^{\qu}}
}}
reproduces the ratio of D-brane tensions, {\it i.e.},
\eqn\ratiotens{
R={T_{p}\o2\pi\rt{\al'} T_{p+1}}=1.
}
They checked this conjecture numerically by using the level truncation.
In section 4, we prove \ratiotens\ analytically.

\newsec{Properties of Neumann Coefficients}
In this section, we summarize some properties of
Neumann coefficients.
To calculate $R$, first we have to know the relation between $M$ and $M'$,
which is summarized in the Appendix B of \RSZclas. 
$M'$ is given by 
\eqn\formMpri{
M'=\lf(\matrix{M'_{00}&M'_{0m}\cr M'_{n0}&M'_{nm}}\ri)=
\lf(\matrix{1-{2\o3}{b\o\V+{b\o2}}&
-{2\o3}{\rt{2b}\o\V+{b\o2}}\bra v_e|\cr
-{2\o3}{\rt{2b}\o\V+{b\o2}}|v_e\ket&
M+{4\o3}{1\o\V+{b\o2}}\big(-|v_e\ket\bra v_e|+|v_o\ket\bra v_o|\big)
}\ri).
}
Following \KOkin, we use a bracket notation, such as $|v_e\ket$, for 
infinite-dimensional vectors with index running from $1$ to $\infty$.
Note that bra and ket are related by the transpose, not by the hermitian 
conjugation.
$|v_e\ket$ and $|v_o\ket$ in \formMpri\ are defined by\foot{
$W_n$ in \RSZclas\ is related to $v_e$ and $v_o$ by $W_n=-\rt{2}(v_e+iv_o)_n$.
}
\eqn\vevodef{
|v_e\ket=E^{-1}|A_e\ket,\quad |v_o\ket=E^{-1}|A_o\ket,
}
where $E_{nm}=\rt{n}\cob_{nm}$ and 
\eqn\defAeo{
(A_e)_n={1+(-1)^n\o2}A_n,\quad (A_o)_n={1-(-1)^n\o2}A_n.
}
$A_n$ is defined by
\eqn\genAn{
\sum_{n={\rm even}}A_nz^n+i\sum_{n={\rm odd}}A_nz^n=
\lf({1+iz\o1-iz}\ri)^{{1\o3}}=\exp\lf({2i\o3}\tan^{-1}z\ri).
}

Another information we need is the spectrum of 
the matrix $M$. This was recently obtained in \spec.
The eigenvector $|k\ket$ of $M$ with eigenvalue $M(k)$  is defined by
\eqn\eigenM{
M|k\ket=M(k)|k\ket.
}
$M(k)$ is given by 
\eqn\Mk{
M(k)=-{1\o2\cosh{\pi k\o2}+1},
}
and $|k\ket$ is given implicitly by the generating function
\eqn\genk{
\bra z|E^{-1}|k\ket={1\o k}(1-e^{-k\tan^{-1}z}),
}
where $\bra z|=(z,z^2,\cdots)$.
The inner product between two eigenvectors is proportional to
the $\cob$-function
\KOkin: 
\eqn\normk{
\bra k|k'\ket=\N(k)\cob(k-k'),
}
where $\N(k)$ is given by
\eqn\Nk{
\N(k)={2\o k}\sinh{\pi k\o2}.
}
We can write down the completeness relation as
\eqn\comp{
{\bf 1}=\int_{-\infty}^{\infty}{dk\o\N(k)}|k\ket\bra k|.
}

We also need to know 
the overlap between $\bra k|$ and  $|v_e\ket,|v_o\ket$.
$\bra k|v_e\ket$ and $\bra k|v_o\ket$ are calculated as
\eqn\kveo{
\bra k|v_e\ket={1\o k}\cdot{\cosh{\pi k\o2}-1\o2\cosh{\pi k\o2}+1},\quad
\bra k|v_o\ket={\rt{3}\o k}\cdot{\sinh{\pi k\o2}\o2\cosh{\pi k\o2}+1}.
}
See Appendix A for the derivation of these equations.

The following identities obtained by Hata and Moriyama\foot{
Note that ${\bf v}_0$, ${\bf v}_1$ and $V_{00}$ in \HataMoriyama\ are
related to our $|v_e\ket$, $|v_o\ket$ and $\V$ by
\eqn\relbfvveo{
{\bf v}_0=-{2\o3}|v_e\ket,\quad {\bf v}_1={2\o{\rt{3}}}|v_o\ket,\quad
V_{00}=\hf\V.
}} \HataMoriyama\ play an essential role in
the calculation of $R$:
\eqn\HMeven{
\bra v_e|{1\o1+3M}|v_e\ket=\qu\V,
}
\eqn\HModd{
\bra v_o|{1\o1-M}|v_o\ket={3\o4}\V.
}
Although $1+3M$ has a kernel $|k=0\ket$ which is twist-odd,
\HMeven\ is well-defined since $1+3M$ is invertible on the twist-even
subspace.

In Appendix B, we present the integral representations 
of the left-hand side of
\HMeven\ and \HModd.

\newsec{Calculation of Ratio}
In this section, we calculate
the ratio $R$ \Ratio\  
using the information in the previous section.
Since $1+3M$ and $1+3M'$  have kernel, 
we have to regularize the ratio of their determinants.
We find that the nontrivial result appears as a twist
anomaly as discussed in \HataMoriyama.  

\subsec{ $\det(1-M')/\det(1-M)$}
Let us first consider the ratio  $\det(1-M')/\det(1-M)$.
From the explicit form of $M'$ in \formMpri, $\det(1-M')$ can be 
written as
\eqn\Mpridet{
\det(1-M')={2\o3}{b\o\V+{b\o2}}
\det\lf(1-M-{4\o3}{1\o\V+{b\o2}}|v_o\ket\bra v_o|\ri).
}
Here we have used the formula
\eqn\detbig{
\det\lf(\matrix{A&B\cr C&D}\ri)=
\det A\det(D-CA^{-1}B).
}
By making use of the formula
\eqn\detrankone{
\det(1+|u\ket\bra v|)=1+\bra v|u\ket
}
and the identity \HModd, 
the ratio $\det(1-M')/\det(1-M)$ turns out to be
\eqn\MprioMdet{\eqalign{
{\det(1-M')\o\det(1-M)}&={2\o3}{b\o\V+{b\o2}}
\det\lf(1-{4\o3}{1\o\V+{b\o2}}{1\o1-M}|v_o\ket\bra v_o|\ri)\cr
&={2\o3}{b\o\V+{b\o2}}
\lf[1-{4\o3}{1\o\V+{b\o2}}\bra v_o|{1\o1-M}|v_o\ket\ri]\cr
&={b^2\o3\lf( \V+{b\o2}\ri)^2}.
}}

\subsec{$\det(1+3M')/\det(1+3M)$}
Next we consider the ratio  $\det(1+3M')/\det(1+3M)$.
Since $1+3M$ and $1+3M'$ have non-zero kernel,
we have to regularize the ratio of their determinants.
The kernel of $1+3M$ is $|k=0\ket$,
and that of $1+3M'$ is given by
\eqn\kerthMpri{
\lf(\matrix{1\cr {8\o\rt{2b}}{1\o 1+3M}|v_e\ket}\ri).
}
We regularize the ratio  $\det(1+3M')/\det(1+3M)$ as 
\eqn\regep{
\lim_{\ep\riya0}{\det(1+3M'+3\ep^2\h{L})\o\det(1+3M+3\ep^2L)}
}
with
\eqn\defhatM{
\h{L}=\lf(\matrix{0&0\cr 0&L}\ri).
}
Here $L=f(M)$ is an arbitrary function of $M$ satisfying
\eqn\Lposi{
L(k)>0,\quad\forall k\in(-\infty,\infty).
}
This condition ensures that the matrix $1+3M+3\ep^2L$ has no kernel.
Note that this regularization \regep\ is
equivalent to the replacement
\eqn\MtoML{
M\riya M+\ep^2L.
}

Using the formula \detbig\ as in the previous subsection, 
the ratio of these determinants is written as
\eqn\thMratiodet{\eqalign{
&{\det(1+3M'+3\ep^2\h{L})\o\det(1+3M+3\ep^2L)}\cr
=&{4\V\o\V+{b\o2}}\det\lf[1+{1\o1+3M+3\ep^2L}\lf(
-{4\o\V}|v_e\ket\bra v_e|+{4\o\V+{b\o2}}|v_o\ket\bra v_o|\ri)\ri].
}}
Since twist-even and twist-odd vectors are 
orthogonal to each other, the following matrix element vanishes:  
\eqn\orthoveo{
\bra v_e|{1\o1+3M+3\ep^2L}|v_o\ket=0.
}
Thus, the matrix inside of the determinant 
in \thMratiodet\
can be rewritten as
\eqn\lastfacthM{\eqalign{
&1+{1\o1+3M+3\ep^2L}\lf(
-{4\o\V}|v_e\ket\bra v_e|+{4\o\V+{b\o2}}|v_o\ket\bra v_o|\ri) \cr
=&\lf(1-{4\o\V}{1\o1+3M+3\ep^2L}|v_e\ket\bra v_e|\ri)
\lf(1+{4\o\V+{b\o2}}{1\o1+3M+3\ep^2L}|v_o\ket\bra v_o|\ri).
}}
Using the formula \detrankone, the ratio \thMratiodet\ 
can be further simplified as 
\eqn\thMinner{\eqalign{
&{\det(1+3M'+3\ep^2\h{L})\o\det(1+3M+3\ep^2L)}\cr
=&{4^2\o\V+{b\o2}}\lf[\qu\V-\bra v_e|{1\o1+3M+3\ep^2L}|v_e\ket\ri]
\lf[1+{4\o\V+{b\o2}}\bra v_o|{1\o1+3M+3\ep^2L}|v_o\ket\ri].
}}
From the identity \HMeven, one can see that the second factor in \thMinner\
is vanishing in the limit $\ep\riya0$.
On the other hand, the last factor in \thMinner\ is divergent
since $1+3M$ is not invertible on the twist-odd subspace.
However, as we will show below, the product of these two  factors
has a well-defined limit when $\ep\riya0$.

From \HMeven, the second factor in \thMinner\ can be rewritten as
\eqn\firstfacre{
\qu\V-\bra v_e|{1\o1+3M+3\ep^2L}|v_e\ket
=\bra v_e|{3\ep^2L\o(1+3M+3\ep^2L)(1+3M)}|v_e\ket.
}
Therefore, \thMinner\ becomes
\eqn\thMepdist{\eqalign{
&{\det(1+3M'+3\ep^2\h{L})\o\det(1+3M+3\ep^2L)}\cr
=&{4^2\o\V+{b\o2}}\bra v_e|{3\ep^2 L\o(1+3M+3\ep^2L)(1+3M)}|v_e\ket
\lf[1+{4\o\V+{b\o2}}\bra v_o|{1\o1+3M+3\ep^2L}|v_o\ket\ri] \cr
=&{4^2\o\V+{b\o2}}\bra v_e|{3\ep L\o(1+3M+3\ep^2L)(1+3M)}|v_e\ket
\lf[\ep+{4\o\V+{b\o2}}\bra v_o|{\ep\o1+3M+3\ep^2L}|v_o\ket\ri].
}}

Now we can take the limit $\ep\riya0$.
For definiteness, we only consider the limit $\ep\riya0^+$, {\it i.e.},
$\ep$ approaches $0$ from above, but we can show that the $\ep\riya0^-$
limit gives the same answer.
Using the formula 
\eqn\deltalim{
\lim_{\ep\riya 0^+}{\ep\o\ep^2+x^2}=\pi\cob(x),
}
we can see that the eigenvalue of the matrix $\ep/(1+3M+3\ep^2L)$
appearing in \thMepdist\ approaches the $\cob$-function of $k$:
\eqn\invthMlim{
\lim_{\ep\riya 0^+}{\ep\o1+3M(k)+3\ep^2L(k)}
={2\o\rt{L(0)}}\cob(k).
}
Therefore, the following type of matrix element
is given by the value at $k=0$: 
\eqn\twistanom{\eqalign{
\lim_{\ep\riya0^+}\bra v|{\ep\o1+3M+3\ep^2L}X|v\ket&=
\lim_{\ep\riya0^+}\int_{-\infty}^{\infty}{dk\o\N(k)}{\ep\o1+3M(k)+3\ep^2L(k)}
X(k)\bra k|v\ket^2 \cr
&=\int_{-\infty}^{\infty}{dk\o\N(k)}{2\o\rt{L(0)}}\cob(k)X(k)\bra k|v\ket^2 \cr
&={2\o\rt{L(0)}}{X(k)\o\N(k)}\bra k|v\ket^2\Big|_{k=0}.
}}
Here we have used the completeness relation \comp\ of $|k\ket$.
This kind of phenomenon that the nontrivial
contribution   comes only from $k=0$ 
is called twist anomaly in \HataMoriyama. 
Substituting the eigenvalue $M(k)$  \Mk\ and
$\bra k|v_{e,o}\ket$  \kveo\
into the general formula \twistanom, and using $\N(0)=\pi$ \Nk,
the limit of the matrix elements appearing in \thMepdist\ can be 
evaluated as  
\eqn\epliminner{\eqalign{
\lim_{\ep\riya0^+}\bra v_e|{3\ep L\o(1+3M+3\ep^2L)(1+3M)}|v_e\ket&
={\pi\o8}\rt{L(0)},\cr
\lim_{\ep\riya0^+}\bra v_o|{\ep\o1+3M+3\ep^2L}|v_o\ket&={\pi\o6}{1\o\rt{L(0)}}.
}}
Finally, the ratio  $\det(1+3M')/\det(1+3M)$ is found to be
\eqn\ratioonethM{
{\det(1+3M')\o\det(1+3M)}\equiv\lim_{\ep\riya0}
{\det(1+3M'+3\ep^2\h{L})\o\det(1+3M+3\ep^2L)}
={4\pi^2\o3\lf(\V+{b\o2}\ri)^2}.
}

\subsec{Proof of $R=1$}
Plugging the ratios \MprioMdet\ and \ratioonethM\
into the definition of $R$ \Ratio,
we arrive at the final result: 
\eqn\Rresult{
R={3\lf(\V+{b\o2}\ri)^2\o\rt{2\pi b^3}}
\lf[{b^2\o3\lf( \V+{b\o2}\ri)^2}\ri]^{{3\o4}}
\lf[{4\pi^2\o3\lf(\V+{b\o2}\ri)^2}\ri]^{\qu}=1.
}
Namely, the ratio of D-brane tensions 
is correctly reproduced from the ratio of the norm of projections!
Note that the result is independent of the parameter $b$
as conjectured in \RSZclas. 

\newsec{Discussions}
In this paper, we showed analytically that the ratio of the norm of
projections correctly reproduces the ratio of D-brane tensions.

Here we would like to discuss on  
the issue of the regularization dependence of $R$.
We showed that the regularization written as a shift of $M$ \MtoML\ 
gives the correct ratio for arbitrary choice of $L$.
Instead of shifting $M$, we can consider some different regularizations,
{\it e.g.}, we can cut-off the dangerous region $k\sim 0$ as
\eqn\intkcut{
\int_{-\infty}^{\infty}dk\riya \int_{|k|>\ep}dk.
} 
Unfortunately, this method leads to a wrong result:
\eqn\cutresu{\eqalign{
\qu\V-\bra v_e|{1\o1+3M}|v_e\ket &\sim 
\qu\V-\int_{|k|>\ep}{dk\o\N(k)}{\bra k|v_e\ket^2\o1+3M(k)}
\sim {\pi\o24}\ep,\cr
\bra v_o|{1\o1+3M}|v_o\ket &\sim  \int_{|k|>\ep}{dk\o\N(k)}
{\bra k|v_o\ket^2\o1+3M(k)}\sim {2\o\pi\ep},\cr
\Longrightarrow {\det(1+3M')\o\det(1+3M)}
 &={4^2\o3\lf(\V+{b\o2}\ri)^2}.
}}  
Therefore, the ratio $R$ seems to be dependent on the regularization.

What is wrong of this regularization and why does our regularization give the
correct result? In the following,
we will present a plausible (but not conclusive) argument
on this issue. 
One obvious bad point of this cut-off regularization is that 
the cut-off procedure does not commute with the matrix multiplication:
\eqn\cutmatpro{
(P_{\ep}XP_{\ep})(P_{\ep}YP_{\ep})\not=P_{\ep}XYP_{\ep},
}
where $P_{\ep}$ is the projection defined by
\eqn\projep{
P_{\ep}=\int_{|k|>\ep}{dk\o\N(k)}|k\ket\bra k|.
}
For example, $(P_{\ep}'M'P_{\ep}')^2\not=P_{\ep}'M'^2P_{\ep}'$ where 
$P_{\ep}'={\rm diag}(1,P_{\ep})$.
In consequence, the associativity of the star product is broken
explicitly under this regularization, and hence
the gauge symmetry of VSFT is not preserved. 
On the other hand, one can show that our regularization \MtoML\
(at least formally)
respects the associativity of the star product if one changes
the Neumann matrices $M^{rs}\riya M_{\ep}^{rs}$ as
\eqn\Mrsep{
M^{11}_{\ep}=M+\ep^2L,\quad 
M^{12}_{\ep}+M^{21}_{\ep}=1-M^{11}_{\ep},\quad 
M^{12}_{\ep}M^{21}_{\ep}=(M^{11}_{\ep})^2-M^{11}_{\ep},
}
and similarly for $M'^{rs}$. 

This is reminiscent of the situation in Yang-Mills
theories. In that case, we have to choose a regularization respecting the 
gauge symmetry, {\it e.g.}, the dimensional regularization.
On the other hand, 
the Pauli-Villars regularization would break the gauge invariance
and lead to a wrong answer.

In our case, the guiding principle of regularization is
the {\it gauge symmetry of VSFT}.
Note that $R$ is a gauge invariant quantity 
since it is written as a ratio of the classical action of lump 
to that of sliver. 
Therefore, one can expect that all the regularizations 
respecting the gauge symmetry will give the
same result, namely $R=1$.



\vskip 8mm
\centerline{{\bf Acknowledgement}}
I would like to thank David Kutasov for discussion.

\appendix{A}{Calculation of $\bra k|v_e\ket$ and $\bra k|v_e\ket$}
In this appendix, we give a derivation of \kveo.
From \genAn, the generating functions of $|A_e\ket$ and $|A_o\ket$
are found to be
\eqn\genfuncAeo{
\bra z|A_e\ket=\cosh\lf({2i\o3}\tan^{-1}z\ri)-1,\quad
\bra z|A_o\ket=-i\sinh\lf({2i\o3}\tan^{-1}z\ri).
}
Let us first consider $\bra k|v_e\ket$.
This can be extracted from the generating function \genfuncAeo\ as
\eqn\brakveth{\eqalign{
\bra k|v_e\ket&=\bra k|E^{-1}|A_e\ket=\int_{-{\pi\o2}}^{{3\pi\o2}}{d\th\o2\pi}
\bra k|E^{-1}|e^{i\th}\ket\bra e^{-i\th}|A_e\ket\cr
&=\int_{-{\pi\o2}}^{{3\pi\o2}}{d\th\o2\pi}{1\o k}(1-e^{-k\tan^{-1}e^{i\th}})
\cosh\lf({2i\o3}\tan^{-1}e^{-i\th}\ri).
}}
Note that the constant term in $\bra z|A_e\ket$ can be neglected
since $\bra k|$ does not have $n=0$ component.
As discussed in \KOkin, this integral can be evaluated by
making a change of integration variable from $\th$ to $x$ as
(see \KOkin\ for detail)
\eqn\thtox{\eqalign{
\tan^{-1}e^{i\th}&={\pi\o4}+ix\qquad (-{\pi\o2}\leq \th \leq{\pi\o2}) \cr
&=-{\pi\o4}-ix\qquad ({\pi\o2}\leq \th \leq{3\pi\o2}).
}}
Then  \brakveth\ is rewritten as
\eqn\kveint{\eqalign{
\bra k|v_e\ket&={2\o\pi k}\int_{-\infty}^{\infty}{dx\o\cosh 2x}
\lf[1-\cosh\lf({\pi k\o4}+ikx\ri)\ri]\cosh\lf({2x\o3}+{\pi i\o6}\ri) \cr
&={2\o\pi k}\int_0^{\infty}{dx\o\cosh 2x}\lf[2\cosh{2x\o3}\cos{\pi\o6}
-\cosh\lf({2x\o3}+ikx\ri)\cosh\lf({\pi k\o4}+{\pi i\o6}\ri)\ri.\cr
&\hskip 55mm \lf.
-\cosh\lf({2x\o3}-ikx\ri)\cosh\lf({\pi k\o4}-{\pi i\o6}\ri)\ri] \cr
&={1\o2k}\lf[2-{\cosh\lf({\pi k\o4}+{\pi i\o6}\ri)
\o\cosh\lf({\pi k\o4}-{\pi i\o6}\ri)}
-{\cosh\lf({\pi k\o4}-{\pi i\o6}\ri)
\o\cosh\lf({\pi k\o4}+{\pi i\o6}\ri)}\ri] \cr
&={1\o k}{\cosh{\pi k\o2}-1\o2\cosh{\pi k\o2}+1}.
}}
Here we have used the formula
\eqn\coshform{
\int_0^{\infty}dx{\cosh ax\o \cosh bx}={\pi\o2b}\cdot{1\o\cos{\pi a\o2b}},\quad
(|{\rm Re}\,a|<{\rm Re}\,b).
}
$\bra k|v_o\ket$ can be calculated in the same way:
\eqn\kvoint{\eqalign{
\bra k|v_o\ket&=-i\int_{-{\pi\o2}}^{{3\pi\o2}}{d\th\o2\pi}
{1\o k}(1-e^{-k\tan^{-1}e^{i\th}})
\sinh\lf({2i\o3}\tan^{-1}e^{-i\th}\ri) \cr
&=-{2i\o\pi k}\int_{-\infty}^{\infty}{dx\o\cosh2x}\sinh\lf({\pi k\o4}+ikx\ri)
\sinh\lf({2x\o3}+{\pi i\o6}\ri) \cr
&=-{2i\o\pi k}\int_0^{\infty}{dx\o\cosh2x}\lf[
\cosh\lf({2x\o3}+ikx\ri)\cosh\lf({\pi k\o4}+{\pi i\o6}\ri)\ri.\cr
&\lf.\hskip 30mm
-\cosh\lf({2x\o3}-ikx\ri)\cosh\lf({\pi k\o4}-{\pi i\o6}\ri)\ri] \cr
&=-{i\o2k}\lf[{\cosh\lf({\pi k\o4}+{\pi i\o6}\ri)
\o\cosh\lf({\pi k\o4}-{\pi i\o6}\ri)}
-{\cosh\lf({\pi k\o4}-{\pi i\o6}\ri)
\o\cosh\lf({\pi k\o4}+{\pi i\o6}\ri)}\ri] \cr
&={\rt{3}\o k}{\sinh{\pi k\o2}\o2\cosh{\pi k\o2}+1}.
}}

\appendix{B}{Integral Representation of Hata-Moriyama's Identities}
Using the completeness relation \comp, 
the left hand sides of \HMeven\ and \HModd\ are written as  
\eqn\veoint{\eqalign{
\bra v_e|{1\o1+3M}|v_e\ket&=\int_{-\infty}^{\infty}{dk\o\N(k)}{1\o1+3M(k)}
\bra k|v_e\ket^2=\qu I,\cr
\bra v_o|{1\o1-M}|v_o\ket&=\int_{-\infty}^{\infty}{dk\o\N(k)}{1\o1-M(k)}
\bra k|v_o\ket^2={3\o4} I,
}}
where $I$ is given by
\eqn\formI{
I=\int_{-\infty}^{\infty}{dk\o k}
{\sinh{\pi k\o2}\o(\cosh{\pi k\o2}+1)(2\cosh{\pi k\o2}+1)} 
=\int_{-\infty}^{\infty}{dt\o t}
{\sinh t\o(\cosh t+1)(2\cosh t+1)}. 
}
Therefore, the identities \HMeven\ and \HModd\ obtained in \HataMoriyama\
are equivalent to
\eqn\Ieqlog{
I=\V=\log{27\o16}.
}

The integral \Ieqlog\ can be evaluated by summing up the residues of 
the following series of  poles on the upper half $t$-plane\foot{
This integral is computed in the recent paper \Mpri\
in a more general setup.} :
\eqn\poles{
t=\pi i+2n\pi i,\quad {2\pi i\o 3}+2n\pi i,\quad 
{4\pi i\o 3}+2n\pi i,\quad (n=0,1,2,\cdots).
}
Then $I$ is written as
\eqn\sumI{
I=\sum_{n=0}^{\infty}\lf(-{4\o2n+1}+{3\o3n+1}+{3\o3n+2}\ri).
}
By introducing the function
\eqn\defIa{\eqalign{
I(a)&=\sum_{n=0}^{\infty}\lf(-{4\o2n+1}+{3\o3n+1}+{3\o3n+2}\ri)a^n\cr
&=\sum_{r=0}^12a^{-\hf}(-1)^r\log(1-(-1)^ra^{\hf})
-\sum_{r=0}^2a^{-{2\o3}}e^{2\pi r i\o3}(1+e^{2\pi ri\o3}a^{{1\o3}})
\log(1-e^{2\pi ri\o3}a^{{1\o3}}),
}}
the summation \sumI\ is evaluated as
\eqn\Ialim{
I=\lim_{a\riya 1}I(a)=\log{27\o16}.
}

\listrefs

\end